\documentclass[apl,twocolumn,showpacs]{revtex4-1}

\usepackage{graphicx}
\usepackage{dcolumn}
\usepackage{bm}
\usepackage{natbib}
\usepackage{amsmath} 
\usepackage{epstopdf}
\usepackage{epsfig}

\draft 

\begin{document}

\title{Opto-Mechano-Fluidic Viscometer}

\author{Kewen Han}
\email{khan56@illinois.edu}
\author{Kaiyuan Zhu}
\author{Gaurav Bahl}
\email{bahl@illinois.edu}
\affiliation{Mechanical Science and Engineering, University of Illinois Urbana Champaign, 1206 W. Green St., Urbana, IL 61801, USA}

\date{\today}

\begin{abstract}
The recent development of opto-mechano-fluidic resonators has provided -- by harnessing photon radiation pressure -- a new microfluidics platform for the optical sensing of fluid density and bulk modulus. 
Here we show that fluid viscosity can also be determined through optomechanical measurement of the vibrational noise spectrum of the resonator mechanical modes.
A linear relationship between the spectral linewidth and root-viscosity is predicted and experimentally verified in the low viscosity regime. 
Our result is a step towards multi-frequency measurement of viscoelasticity of arbitrary fluids, without sample contamination, using highly sensitive optomechanics techniques. 
\end{abstract}

\maketitle 

The optical and mechanical modes of high-Q resonant systems can be parametrically coupled through optical radiation-pressure induced mechanical instabilities \cite{Carmon2005, Rokhsari05, Kippenberg2005}. 
Ultra-sensitive optomechanical sensors of acceleration \cite{Krause2012, Hutchison2012}, mass \cite{Liu:iu, Liu2013}, and forces \cite{Liu2012,Gavartin2012} have been demonstrated based on this concept.
While these previous optomechanics demonstrations have focused on solid state sensors, recent demonstrations of opto-mechano-fluidic resonators (OMFRs) have enabled applications with liquids \cite{Bahl2013a, HyunKim2013} and gases \cite{Han2014}. 
Confinement of a liquid inside the resonator, as opposed to outside it, prevents acoustic energy from leaking out of the resonator \cite{Burg2007a}. It has been shown that mechanical modes in OMFRs are able to penetrate into the fluids, enabling sensing of the fluidic environment within \cite{Bahl2013a, HyunKim2013}.
In addition, OMFR operational frequencies extend from a few MHz to the 11 GHz regime, presenting an opportunity to study dynamics over broad timescales. Especially when the frequency is high, the viscoelastic nature of the fluid becomes prominent \cite{Martin1989}.
OMFRs can thus provide a novel path towards high-resolution analysis of the viscoelastic properties of fluids and bioanalytes. 
In this work, we demonstrate the first experimental system for optomechanically measuring the dynamic viscosity, $\mu$, of various test fluids using sample volumes in the nanoliter regime.

\begin{figure}[htbp]
	\centering
	\includegraphics[width=0.42\textwidth, clip=true, trim=1.4in 2.0in 2.6in 4.2in]{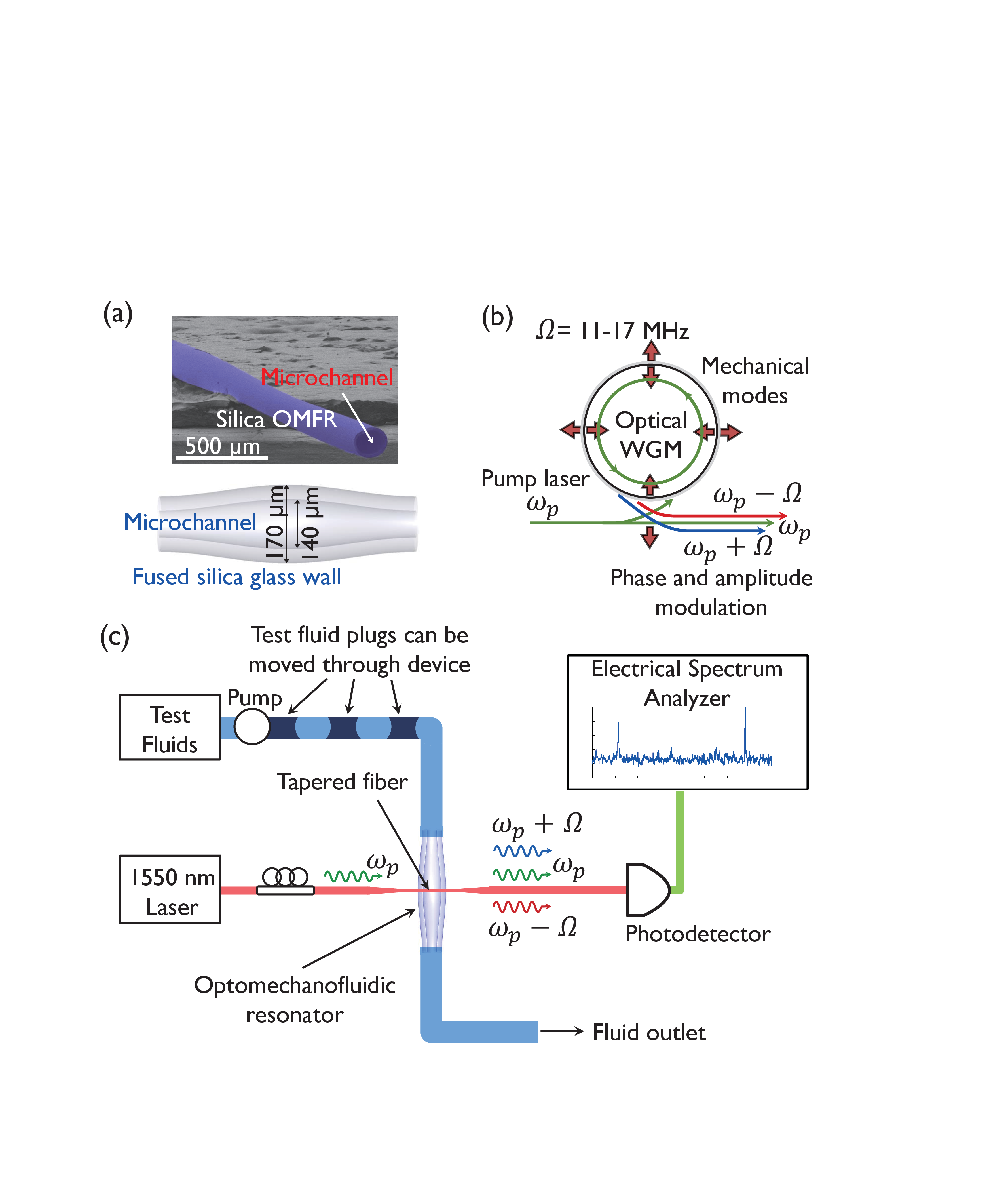}
	\caption{{\bf Experimental overview:}
	(a) SEM of a fused-silica opto-mechano-fluidic resonator (OMFR). The device has a 170 $\mu$m diameter and a 15 $\mu$m wall thickness  at the widest point. 
	(b) Light of frequency $\omega_{p}$ is coupled to ultra-high-Q optical whispering gallery modes (WGMs) through a tapered optical fiber. Modulation of the device geometry at mechanical frequency $\Omega$ generates both upper and lower optical sidebands of the pump light at $\omega _p \pm  \Omega$ \cite{Schliesser2010}. 
	(c) The temporal interference of optical signals at the photodetector generates an electronically measurable signal, allowing the mechanical power spectrum to be measured using an electrical spectrum analyzer. Test fluids can be pumped in and out of the resonator, changing the effective mass and stiffness as well as the damping loss rate.}
	\label{fig:overview} 
\end{figure}

The fabrication of OMFRs (Fig. \ref{fig:overview}(a)) has been reported in \cite{Bahl2013a} and is based on previously established techniques in optofluidics \cite{Lacey2007}. Briefly, fused-silica capillary preforms are heated and softened by means of high-power CO$_2$ lasers and drawn linearly into microcapillaries. Modulation of the laser power varies the capillary diameter creating “bottle”-shaped OMFRs. In this work, the device we use has a $\sim$ 170 $\mu$m diameter and $\sim$ 15 $\mu$m wall thickness. One end of the device is left open while the other end is connected to a syringe through which analytes can be infused (Fig. \ref{fig:overview}(c)). The sensing volume contained within (Fig. \ref{fig:overview}(a)) is about 20 nl, and can be reduced by changing the fabrication parameters.

\begin{figure*}[htbp]
	\centering
	\includegraphics[width=0.7\textwidth, clip=true, trim=0.in 11in 0.5in 0.1in]{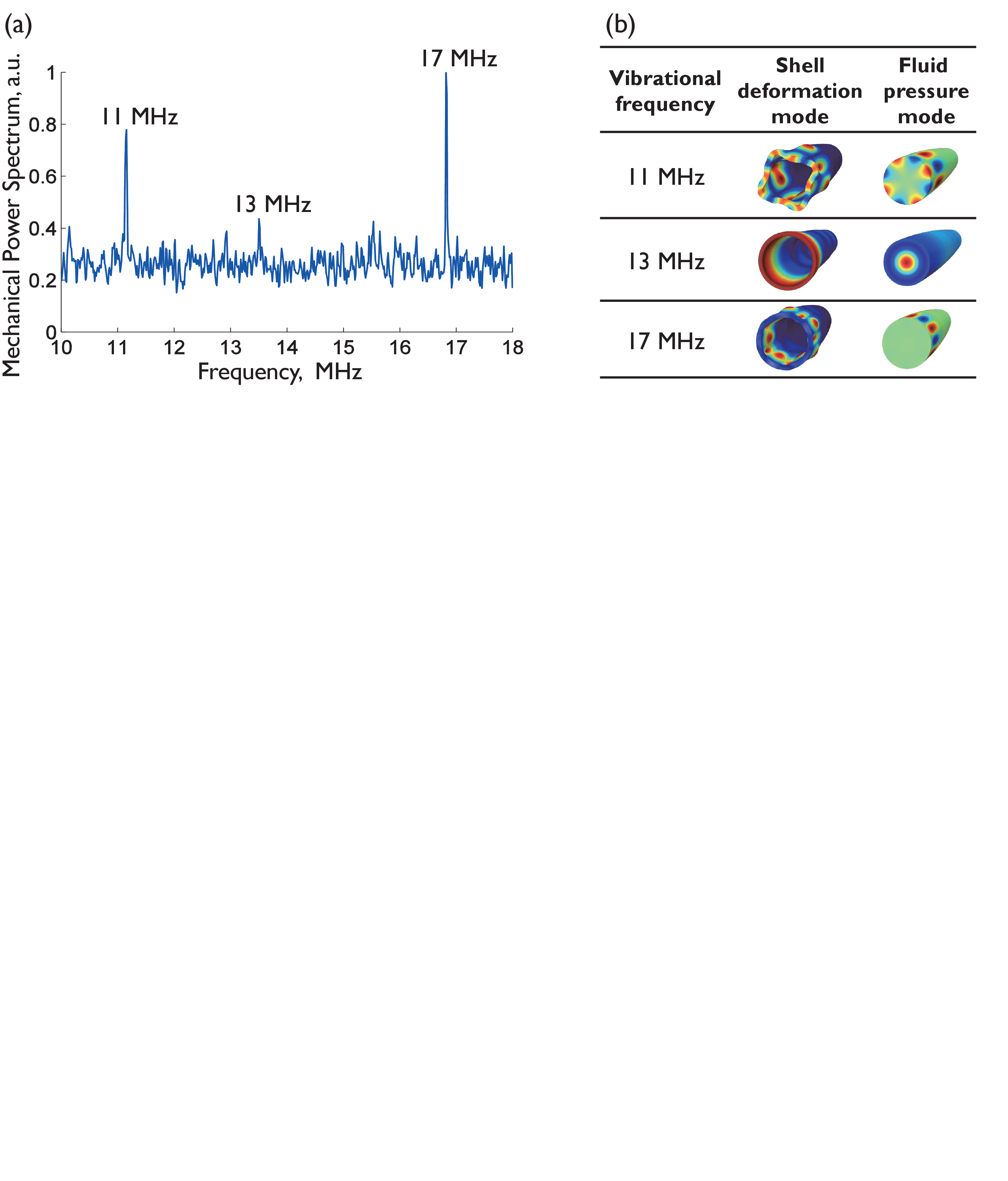}
	\caption{(a) Mechanical power spectrum using above-threshold, continuous-wave laser excitation power shows three mechanical modes. (b) Multiphysical simulations of solid OMFR shell and coupled pressure waves in fluid for the 11 MHz (high order wineglass mode), 13 MHz (breathing mode), and 17 MHz (high order wineglass mode) vibrational resonances.} 
	\label{fig:modeshape}
\end{figure*}

\begin{figure*}[t]
	\centering
	\includegraphics[width=0.78\textwidth, clip=true, trim=0.3in 4.95in 0.2in 2.2in]{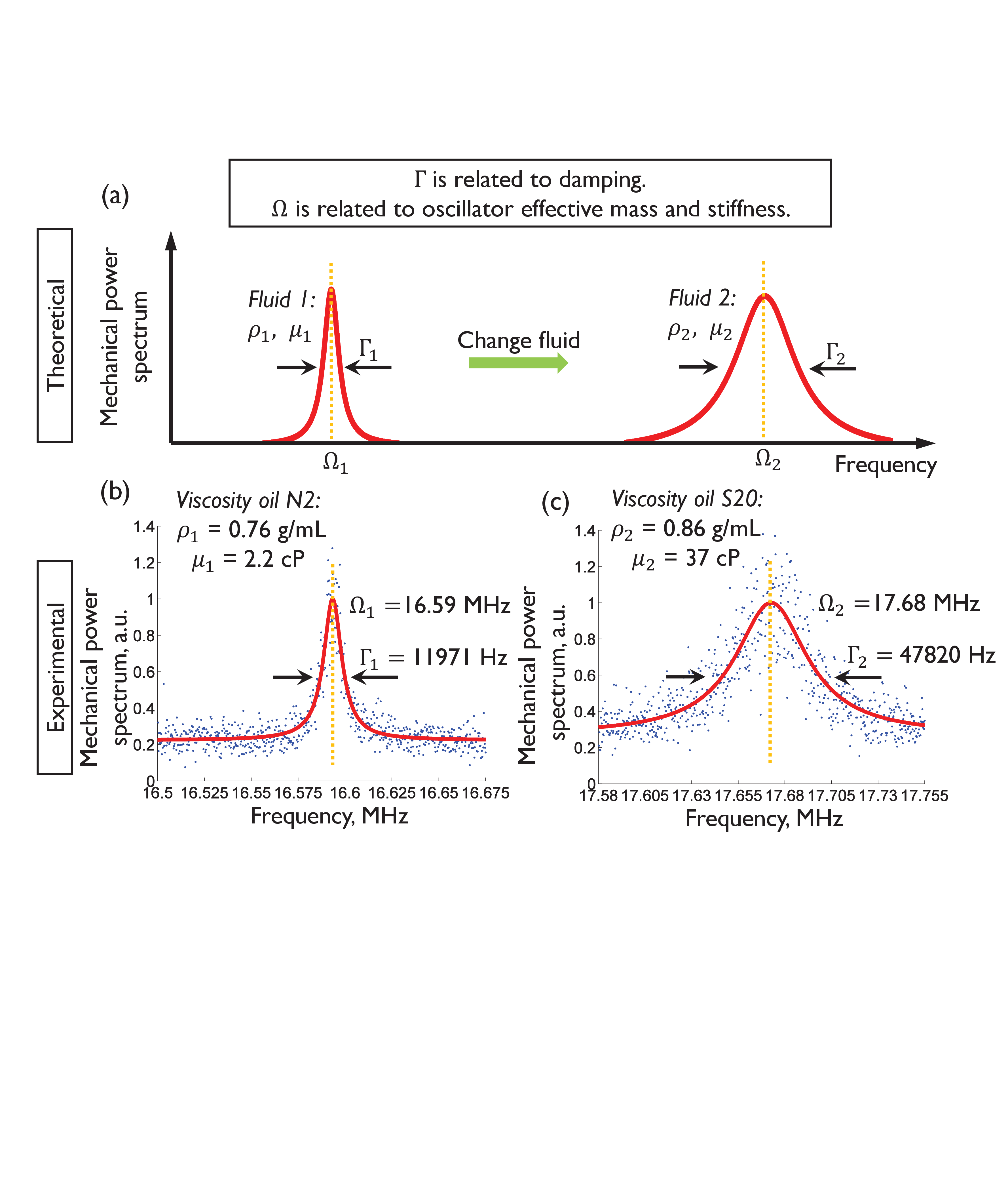}
	\caption{(a) Theoretically, the vibrational noise spectrum contains information about both the mechanical damping rate and the effective mass and stiffness of the hybrid system. The linewidth (-3 dB bandwidth) of the spectrum, $\Gamma$, is related to damping; and the vibrational frequency, $\Omega$, is related to the oscillator effective mass and stiffness.  Example measurement of the vibrational noise spectrum of $\sim$ 17 MHz mode with (b) N2 viscosity oil, and with (c) S20 viscosity oil. In both cases, there are linewidth and center frequency changes.} 
	\label{fig:measurement}
\end{figure*}

Continuous-wave 1550 nm laser light is coupled into the optical whispering-gallery modes (Q-factor $\sim$ 10$^7$) of the OMFR by evanescent coupling \cite{Knight1997} through a tapered optical fiber (Fig. \ref{fig:overview}(c)). The taper is not in contact with the device, which prevents additional damping effects.
The radiation pressure of light is capable of actuating eigenmechanical oscillations through the optomechanical parametric instability \cite{HyunKim2013} in this device (Fig. \ref{fig:modeshape}(a)). Mechanical modulation of the device geometry generates optical sidebands of the input light (Fig. \ref{fig:overview}(b)). Even when the parametric actuation threshold power is not reached, stochastic thermal fluctuations (Langevin noise force) provide a detectable amount of quiescent energy to the mechanical degrees of freedom. We can electronically measure the noise spectrum of the mechanical mode by observing the beating between input and scattered light on a photodetector (Fig. \ref{fig:overview}(c)). 
In this work, $\Omega \approx$ 11 MHz, 13 MHz, and 17 MHz vibrational modes are selected (Fig. \ref{fig:modeshape}(a)). According to computational models, these modes are high-order wineglass modes and a breathing mode, where both fluid and shell are coupled into a hybrid eigenmode (Fig. \ref{fig:modeshape}(b)).
For a continuously driven system, the mechanical damping losses in these hybrid shell-fluid modes can be obtained through optical measurement of the linewidth of the Lorentzian shaped  mechanical noise spectrum as described in Fig. \ref{fig:measurement}. 
Information on both the mechanical damping rate and the effective mass and stiffness of the hybrid system are embodied in the vibrational noise spectrum. By measuring the linewidth of the spectrum, $\Gamma$, we can quantify the mechanical damping rate of the system; by measuring the center frequency of the spectrum, $\Omega$, we can quantify the oscillator effective mass and stiffness.

\begin{figure}[tbp]
	\centering
	\includegraphics[width=0.45\textwidth, clip=true, trim=2.8in 5.2in 5in 1.9in]{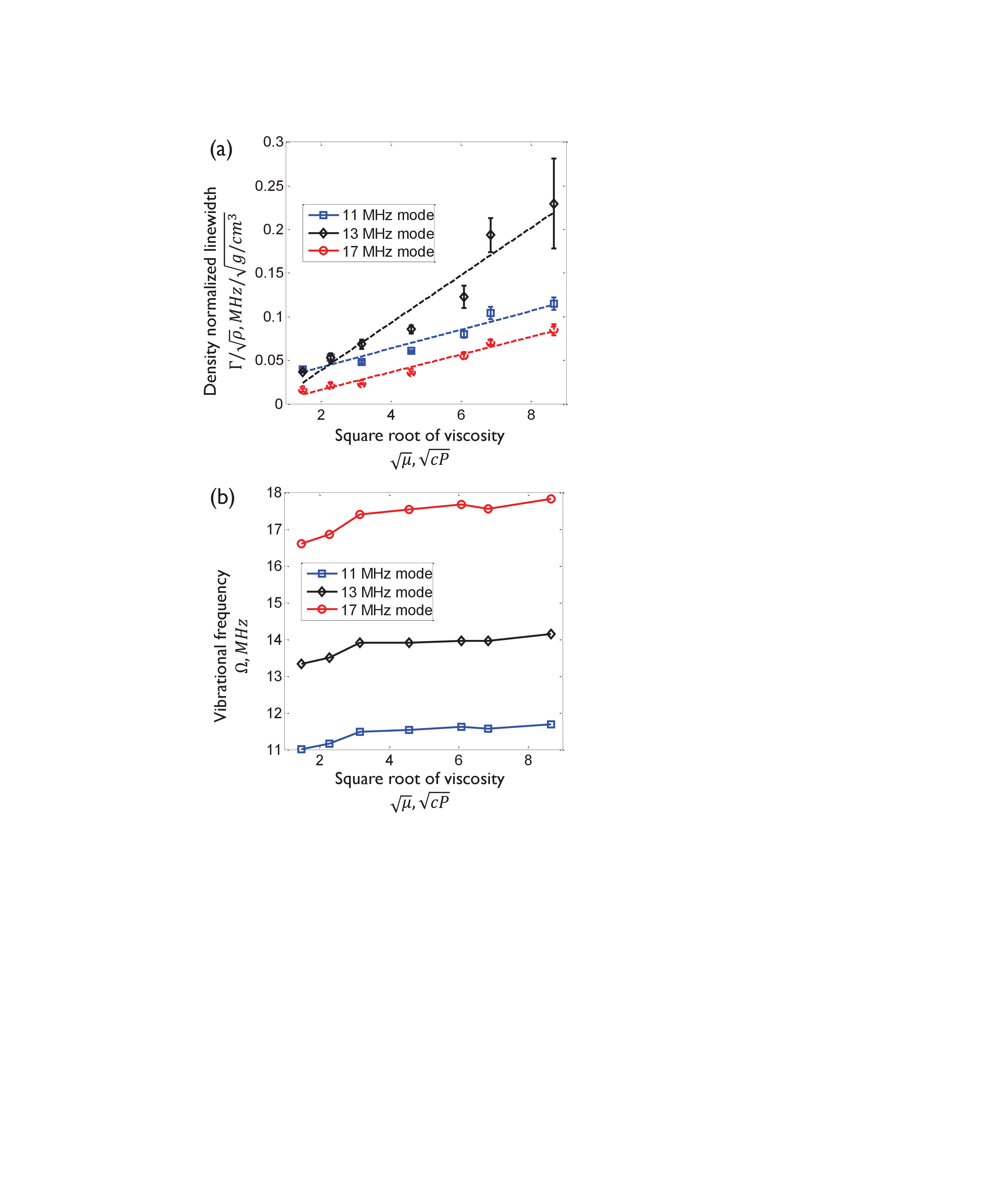}
	\caption{(a) Measured linewidth of selected mechancial modes operating with viscosity standard oils shows that the damping loss rate $\Gamma$ increases linearly with square root of viscosity $\sqrt{\mu}$. Slope differences between different modes indicate different modeshapes (Fig. \ref{fig:modeshape}(b)). Dashed lines are the linear fits. (b) Measured frequency of selected mechanical modes operating with viscosity standard oils. The three selected modes have similar frequency trend. This increasing frequency trend is likely caused by variation in density and the speed of sound of the test fluids \cite{HyunKim2013}.} 
	\label{fig:results}
\end{figure}

Optomechanical self oscillation \cite{Bahl2012a, HyunKim2013}, however, narrows the vibrational linewidth due to amplification, which affects the ability to measure intrinsic oscillator damping. Here we make sure to employ subthreshold input optical power to avoid amplifying the mechanical motion to the self oscillation point. This preserves the intrinsic damping and natural linewidth of the vibrational spectrum, allowing us to quantify intrinsic loss rates. 
To calibrate the optomechanical viscometers, we use seven viscosity standard oils (Cannon Instrument Company -- Table \ref{tab:oils}). 
For each of the three vibrational modes in Fig. \ref{fig:modeshape}(a), we plot in Fig. \ref{fig:results}(a) the density normalized experimentally measured mechanical mode linewidth $\Gamma / \sqrt{\rho}$, against the square root of the viscosity, $\sqrt{\mu}$.

\begin{table}[t]
  \centering
  \caption{Properties of the calibration viscosity oils}
    \begin{tabular}{ccc}
    \hline
    \textbf{Sample} & \textbf{Density (g/mL)} & \textbf{Viscosity (cP)} \\
    \hline
    
    \textbf{N2} & 0.762 & 2.2 \\
    \hline
    \textbf{N4} & 0.787 & 5.2 \\
    \hline
    \textbf{S6} & 0.878 & 10 \\
    \hline
    \textbf{N10} & 0.884 & 21 \\
    \hline
    \textbf{S20} & 0.863 & 37 \\
    \hline
    \textbf{N26} & 0.820  & 47 \\
    \hline
    \textbf{N35} & 0.868 & 75 \\
    \hline
    \end{tabular}%
  \label{tab:oils}%
\end{table}%

The results in Fig. \ref{fig:results}(a) can be understood by considering the nature of viscous damping in OMFR. The geometry of the resonator is a shell, which locally resembles the thin plate case discussed in \cite{Martin1989}.
Because of liquid entrainment within the resonator, viscous damping, associated with both shear and normal motion of the fluid relative to the resonator wall, occurs near the solid-fluid interface. For thin shells or plates, this damping is dominated by the shear motion of the fluid relative to the resonator wall.
The attenuation rate due to viscous damping is proportional to $\sqrt{ \rho \mu }$ at low values of viscosity \cite{Martin1989}. At high viscosity, attenuation saturates due to the viscoelastic nature of the fluid.  Using Maxwell's model of a viscoelastic fluid, a critical viscosity separating low and high viscosity regimes can be defined by $\mu_{c}= \tau G_{\infty}$ at which $2 \pi \Omega \tau=1$, where $\Omega$ is the mechanical vibrational frequency, $ \tau $ is the viscoelastic relaxation time in the liquid, and $ G_{\infty} $ is the high-frequency elastic rigidity modulus. Assuming a typical $G_{\infty}$ value of 1 GPa \cite{Martin1989}, the $ \mu_{c} $ for a 10 MHz device is $ 1.6 \times 10^{4} $ cP, indicating that our experiment is well within the linear regime.
Since the intrinsic losses of the mechanical modes without liquids are very low (Q$_m$ $\sim$ 10$^3$-10$^4$ without fluids), the acoustic energy loss primarily arises from viscous damping associated with the fluid (Q$_m$ $\sim$ 10$^1$-10$^3$ with fluids).
By electronically measuring the vibrational noise spectrum of the mechanical mode, the mechanical mode linewidth, $\Gamma$, is obtained (Fig. \ref{fig:measurement}(b),(c)).
Since $\Gamma$ is proportional to loss rate, it is also proportional to $\sqrt{\rho \mu}$ at low values of viscosity. In order to isolate the effects of viscosity, we normalize the linewidth against $\sqrt{\rho}$ to obtain Fig. \ref{fig:results}(a). We note that the linewidth slopes of  the 11 MHz and 17 MHz are similar to but lower than that of the 13 MHz mode, potentially indicating difference in the mechanical mode families.

In addition, the measured relationship between the line center frequency $\Omega$ and $\sqrt{\mu}$ is plotted in Fig. \ref{fig:results}(b). We see that the frequencies of all three modes are not constant but follow the same increasing trend. We believe that the frequency shift is correlated with the density and speed of sound change in the test fluids as has been established previously \cite{HyunKim2013}.

We now proceed to analyze how the sensitivity and dynamic range are influenced by the OMFR shell thickness. As discussed in \cite{Martin1989}, the attenuation rate is inversely proportional to the plate thickness due to the fact that a thicker plate can transmit more wave energy compared with the amount lost in viscous dissipation. Thus, in the OMFR case, a thinner shell should be used to increase the viscosity sensitivity. In contrast, a thicker shell should be used to expand the viscosity measurement dynamic range. However, as revealed by \cite{Bahl2012}, thick shells can prevent the acoustic excitation from interacting with the liquid and a suitable balance must be sought. In contrast, we note that thin-shell resonators are also subject to potentially undesirable pressure effects \cite{Han2014} generated by pumped liquids.

Recent technological advances in optofluidics \cite{Fan2011} have enabled biochemical sensors that employ many different optical techniques – such as refractive index measurement \cite{White2006,Arnold2009,Yang2014}, fluorescence  \cite{Ganesh2007}, and surface enhanced Raman spectroscopy \cite{Nie1997}. The recent introduction of optically-interfaced acoustics into optofluidic devices \cite{Bahl2013a, HyunKim2013}, inspired by lab-on-a-chip mechanical sensors \cite{Burg2007a}, is providing researchers with a new mechanical degree of freedom by which to perform such biochemical analyses. In this work we have invoked these opto-mechano-fluidic techniques to develop the first microfluidic optomechanical sensor of liquid viscosity.
Our result supplements established passive \cite{Mason1995,Waigh2005} and active \cite{Waigh2005,Pelton2013} microrheological techniques for the optical measurement of fluid viscoelasticity, but does not contaminate the fluid with dispersed particles.
Another advantage of our all-silica fiber-interfaced device is the operability in high temperature, remote, and electromagnetically noisy environments, such as engines, oil wells, and reaction chambers. Finally, the GHz-regime multifrequency capability of OMFRs \cite{Bahl2013a} can enable high-frequency probes for mapping viscoelastic properties of fluids and boundary layers. In the long-term, we envision high-throughput viscoelastic measurements on flowing living cells, essentially enabling a novel acoustic flow cytometry.

\begin{acknowledgments}
Funding for this research was provided through a University of Illinois Startup Grant. We would like to acknowledge stimulating discussions and guidance from Prof. Rashid Bashir, Prof. Randy Ewoldt, Prof. Taher Saif, and Prof. David Saintillan.
\end{acknowledgments}

\end{document}